\documentclass[nolineno]{aastex631}

\usepackage{natbib}
\usepackage{amsmath}
\usepackage{hyperref}  
\usepackage{nameref}
\usepackage{babel}

\graphicspath{{./}{figures/}}

\received{}
\revised{} 
\accepted{} 
\submitjournal{}

\shorttitle{He-accreting ONe WDs and AIC}
\shortauthors{Zhang et al.}

\begin{document}
\title{He-accreting oxygen-neon white dwarfs and accretion-induced collapse events}

\email{wuchengyuan@ynao.ac.cn}
\email{wangbo@ynao.ac.cn}

\author[0009-0006-8370-3108]{Zhengyang Zhang}
\affiliation{Yunnan Observatories, Chinese Academy of Sciences, Kunming 650011, PR China}
\affiliation{University of the Chinese Academy of Sciences, 19A Yuquan Road, Shijingshan District, Beijing 100049, PR China}
\affiliation{International Centre of Supernovae, Yunnan Key Laboratory, Kunming, 650216, China}

\author{Chengyuan Wu}
\affiliation{Yunnan Observatories, Chinese Academy of Sciences, Kunming 650011, PR China}
\affiliation{International Centre of Supernovae, Yunnan Key Laboratory, Kunming, 650216, China}

\author{Amar Aryan}
\affiliation{Aryabhatta Research Institute of Observational Sciences (ARIES), Manora Peak, Nainital-263001, India}
\affiliation{Graduate Institute of Astronomy, National Central University, 300 Jhongda Road, 32001 Jhongli, Taiwan}

\author{Shuai Zha}
\affiliation{Yunnan Observatories, Chinese Academy of Sciences, Kunming 650011, PR China}
\affiliation{International Centre of Supernovae, Yunnan Key Laboratory, Kunming, 650216, China}

\author{Dongdong Liu}
\affiliation{Yunnan Observatories, Chinese Academy of Sciences, Kunming 650011, PR China}
\affiliation{International Centre of Supernovae, Yunnan Key Laboratory, Kunming, 650216, China}

\author{Bo Wang}
\affiliation{Yunnan Observatories, Chinese Academy of Sciences, Kunming 650011, PR China}
\affiliation{International Centre of Supernovae, Yunnan Key Laboratory, Kunming, 650216, China}



\begin{abstract}
It has been widely accepted that mass-accreting white dwarfs (WDs) are the progenitors of Type Ia supernovae or electron-capture supernovae. Previous work has shown that the accretion rate could affect the elemental abundance on the outer layers of CO WDs, and therefore affect the observational characteristics after they exploded as SNe Ia. However, it has not been well studied how elemental abundance changes on the outer layers of He-accreting ONe WDs as they approach the Chandrasekhar mass limit. In this paper, we investigated the evolution of He-accreting ONe WDs with MESA. We found that a CO-rich mantle will accumulate beneath the He layers resulting from the He burning, after which the ignition of the CO-rich mantle could transform carbon into silicon (Si). The amount of Si produced by carbon burning is strongly anti-correlated with the accretion rate. As the ONe WD nearly approaches the Chandrasekhar mass limit ($\rm M_{ch}$) through accretion, it is likely to undergo accretion-induced collapse (AIC), resulting in the formation of the neutron star (NS).

\end{abstract}

\keywords{White dwarf stars (1799), Stellar evolution (1599), Supernovae (1668)}


\section{Introduction}\label{sec1}

White dwarfs (WDs) represent the final evolutionary stage for single stars with initial masses below $8$-$10{\rm M}_{\odot}$. They are typically classified into three categories based on elemental composition: helium (He) WDs, carbon-oxygen (CO) WDs, and oxygen-neon (ONe) WDs \citep[e.g.][]{1970AcA....20...47P, 1994MNRAS.270..121H}. If the WD in a binary system can accrete material from its non-degenerate companion star, then the WD is known as mass-accreting WD \citep[e.g.][]{2004A&A...419..623Y}. These mass-accreting WDs play a crucial role in the evolution of various astronomical phenomena, including novae, Type $\mathrm{I}$a supernovae, and accretion-induced collapse (AIC) events \citep[e.g.][]{1991ApJ...367L..19N}. 

CO WD that accrete He-rich material is widely regarded as one of the promising channels for producing Type $\mathrm{I}$a supernovae \citep[e.g.][]{2009ApJ...701.1540W, 2009MNRAS.395..847W, 2009ApJ...699.2026R, 2010AN....331..218P, 2012NewAR..56..122W}. In this scenario, the mass accretion rate plays a crucial role in determining the evolutionary fate of the WD. Under a high mass accretion rate, CO WD will evolve to He giants. Conversely, if the mass accretion rate is too low, CO WD will experience multiple He flashes, for which the WD cannot accumulate all the accreted material \citep[e.g.][]{2014MNRAS.445.3239P}. Therefore, the steady He layer burning only occurs in a narrow range, in which the accreted He can be completely burned into carbon and oxygen \citep[e.g.][]{1991ApJ...367L..19N, 2015A&A...584A..37W, 2016RAA....16..160W, 2018ApJ...863..125K}. Importantly, within this stable burning regime, a critical accretion rate exists. Beyond this rate, off-center carbon ignition occurs during He accretion \citep[e.g.][]{2017MNRAS.472.1593W}. Recent work claimed that the C burning temperature is higher than previous predictions if the mass accretion rate exceeds the critical rate. In this situation, neon is exhausted immediately once it appears, leading to the formation of a silicon-rich mantle. Consequently, the final fate of the WD in this situation may either be OSi WD if its final mass is lower than Chandrasekhar mass limit (${\rm M}_{\rm ch}$) or be iron-core-collapse supernova (Fe-CCSN, e.g. \citealt{2019MNRAS.486.2977W}). 

Theoretically, the outcome of mass-accreting ONe WD differs from that of CO WD. The electron capture processes of $^{\rm 24}{\rm Mg}$ and $^{\rm 20}{\rm Ne}$ will occur when the mass of ONe WD approaches ${\rm M}_{\rm ch}$, resulting in the formation of electron-capture supernovae (ECSNe). The process by which a mass-accreting ONe WD collapses to a neutron star (NS) is called accretion-induced collapse (AIC) \citep[][]{1980PASJ...32..303M, 1984ApJ...277..791N, 2016ApJ...830L..38M}. Up to now,  AIC is a plausible phenomenon without unambiguous identification. Post AIC events can be identified as low/intermediate-mass X-ray binaries, and subsequently as low/intermediate-mass binary pulsars. If the companion star has a very low mass, an ultracompact X-ray binary (UCXB) may also form \citep[e.g.][]{2023MNRAS.521.6053L}. The energy released in these weak explosions is estimated at approximately $3\times10^{49}\, \rm erg$. However, when magnetic fields are considered, magnetic stresses can result in explosions with energies exceeding $10^{50}\, \rm erg$. {Furthermore, the presence of a magnetic field facilitates stable mass growth through accretion. This leads to potential AIC progenitor systems having shorter initial orbital periods and lower donor masses \citep[e.g.][]{2022MNRAS.509.6061A}.

The mass of the ejecta, both with and without the influence of magnetic fields, is predicted to vary from $0.1$ to $10^{-3}\, \rm M_{\odot}$ \citep[e.g.][]{2006ApJ...644.1063D, 2007ApJ...669..585D}. The amount of synthesized nickel in these explosions is also relatively small, typically around $10^{-3} \, \rm M_{\odot}$ \citep[e.g.][]{1999ApJ...516..892F, 2009ApJ...695..208W}. \citet{2009MNRAS.396.1659M} suggested an interesting scenario, where, if the WD spins rapidly before collapse, a centrifugally-supported disk forms around the newly formed NS. The disk expands to larger radii due to the accretion, which undergoes a recombination of free nuclei into helium. This process releases enough energy to unbind what remains in the disk. As a result, a disk wind could potentially eject approximately $10^{-2}\, \rm M_{\odot}$ of nickel. Following this exploration, \citet{2010MNRAS.409..846D} employed a radiation transfer code to compute the light curve and spectrum of the ejecta after disk disruption. For ejecta masses of $10^{-2}\, \rm M_{\odot}$ and $3\times 10^{-3}\, \rm M_{\odot}$, the light curve achieves its peak luminosity in less than one day, with values of $2\times10^{41} \rm erg\, s^{-1}$ and $5\times10^{40} \rm erg\, s^{-1}$, respectively. These luminosities are fainter by a factor of $\sim$ 100 compared to a typical SN Ia. Near the peak luminosity, the spectrum is predominantly characterized by a Doppler-spread nickel feature, without significant spectral lines. However, between 3 and 5 days after the explosion, distinct calcium lines become apparent in the infrared band. Despite the intrinsic faintness of AIC events, future facilities like the Vera C. Rubin Observatory significantly boost our capacity to detect faint astronomical sources \citep[e.g.][]{2019ApJ...873..111I}. 

There are two classical progenitor models for AIC events: the single degenerate model and the double degenerate model \citep[e.g.][]{2020RAA....20..135W}. In the single degenerate AIC model, the potential companions may be main sequence stars, red giants, or helium stars \citep[e.g.][]{2013A&A...558A..39T, 2022MNRAS.510.6011W}.  The AIC explosion is expected to collide with its companion star, stripping between $10^{-4}$ and $10^{-2}\, \rm M_{\odot}$ of the matter from the companion. Specifically, if the companion is a red giant, the initial emission would be characterized by X-rays lasting approximately an hour. This would be followed by a more prolonged visible transient, with a peak absolute magnitude ranging from -16 to -18. Afterward, the light curves of the transient are powered by shock cooling, which would last for several days to a week \citep[e.g.][]{2014ApJ...794...28P}.

Although previous works have already investigated the evolution of ONe WDs accreting helium-rich material \citep[e.g.][]{2017ApJ...843..151B}, the process of carbon mantle ignition in ONe WDs remains uncertain (under high accretion rate). In this work, we aim to simulate the evolution of He-accreting ONe WDs up to their near approach to the $\rm M_{ch}$ and to speculate on their ultimate fate. The structure of this paper is as follows: In Section \ref{sec2}, we present the detailed methods and results of our simulations of He-accreting ONe WD evolution. In Section \ref{sec3}, we provide a comprehensive discussion of our results. Finally, we summarize our findings in Section \ref{sec4}.

\section{Progenitor}\label{sec2}

\subsection{Methods}\label{sec2.1}

To simulate the long-term evolution of ONe WDs accreting helium-rich materials, we utilize the advanced stellar evolution software, Modules for Experiments in Stellar Astrophysics (MESA version 10398) \citep[e.g.][]{2011ApJS..192....3P, 2013ApJS..208....4P, 2015ApJS..220...15P, 2018ApJS..234...34P, 2019ApJS..243...10P}. In our simulations, the initial masses of ONe WDs are in the range of 1.2 to 1.3 $\rm M_{\odot}$, respectively. The composition of ONe WDs is 50\% of $^{16}\rm O$, 45\% of $^{20}\rm Ne$, and 5\% of $^{24}\rm Mg$ \citep[e.g.][]{2018RAA....18...36W}. The elemental abundances are the same as adopted in \citet{2015MNRAS.453.1910S} and are similar to the abundances of ONe core which is evolved from the intermediate-mass stars by \citet{2013ApJ...771...28T}. The He-rich material accreted onto the ONe WDs consists of 98\% He and Z=0.02. Our primary purpose was to investigate the dependence of mass accretion rates and WD masses on the production of the silicon-rich mantle. Therefore, we designed two sets of models: 1. Using a 1.2$\rm M_{\odot}$ ONe WDs with a range of accretion rates ($2$-$5 \times 10^{-6} \,{\rm M}_{\odot} \, {\rm yr^{-1}}$); 2. Using different WD masses(1.2 to 1.3 $\rm M_{\odot}$) with the same mass accretion rate($4 \times 10^{-6} \,{\rm M}_{\odot} \, {\rm yr^{-1}}$). We incorporate both ion and electron Coulomb corrections into our simulations, as described in previous studies \citep[e.g.][]{2009PhRvE..79a6411P, Itoh_2002}. In MESA, the "neu" module calculates the energy loss rates from neutrinos and their derivatives. These neutrinos are produced by various processes, including plasma decay, electron-positron annihilation, bremsstrahlung, recombination, and photoneutrino processes \citep{2011ApJS..192....3P}. Weak reaction rates are considered in MESA by setting:
\\
\begin{center}
{\small use\_special\_weak\_rates =.true.}\\
{\small special\_weak\_states\_file = 'weak.states'}\\
{\small special\_weak\_transitions\_file = 'weak.transitions'}\\
\end{center}
These settings calculate the weak interaction processes of electron capture and beta decay using the formulation of \citet{2015MNRAS.453.1910S}. The nuclear reaction networks include {\small wd\_aic.net} and {\small co\_burn.net}, in which nuclear reactions of He, C, O, Ne burning as well as electron capture reactions of $^{\rm 24}{\rm Mg}$ and $^{\rm 20}{\rm Ne}$ are considered. To avoid numerical errors, we employ the super-Eddington wind assumption as the mechanism for mass loss during the helium flash and carbon flash \citep[e.g.][]{2013ApJ...762....8D, 2013ApJ...778L..32M, 2015A&A...584A..37W, 2017A&A...604A..31W, 2017gacv.workE..56K, 2018RAA....18...36W, 2019MNRAS.483..263W, 2020MNRAS.495.1445W}. The Ledoux criterion is used as the criterion for convective instability in all models. We adopted a value of 0.014 for the convective overshoot parameter.

\subsection{Results} \label{sec2.2}

We consider the evolution of a 1.2$ \, \rm M_{\odot}$ ONe WD accreting helium-rich material at an accretion rate of $3 \times 10^{-6} \, \rm M_{\odot} \, yr^{-1}$ as a representative example. (As shown in Figure~\ref{fig1} to Figure~\ref{fig3})

\begin{figure}
    \centering
    \resizebox{1\hsize}{!}{\includegraphics{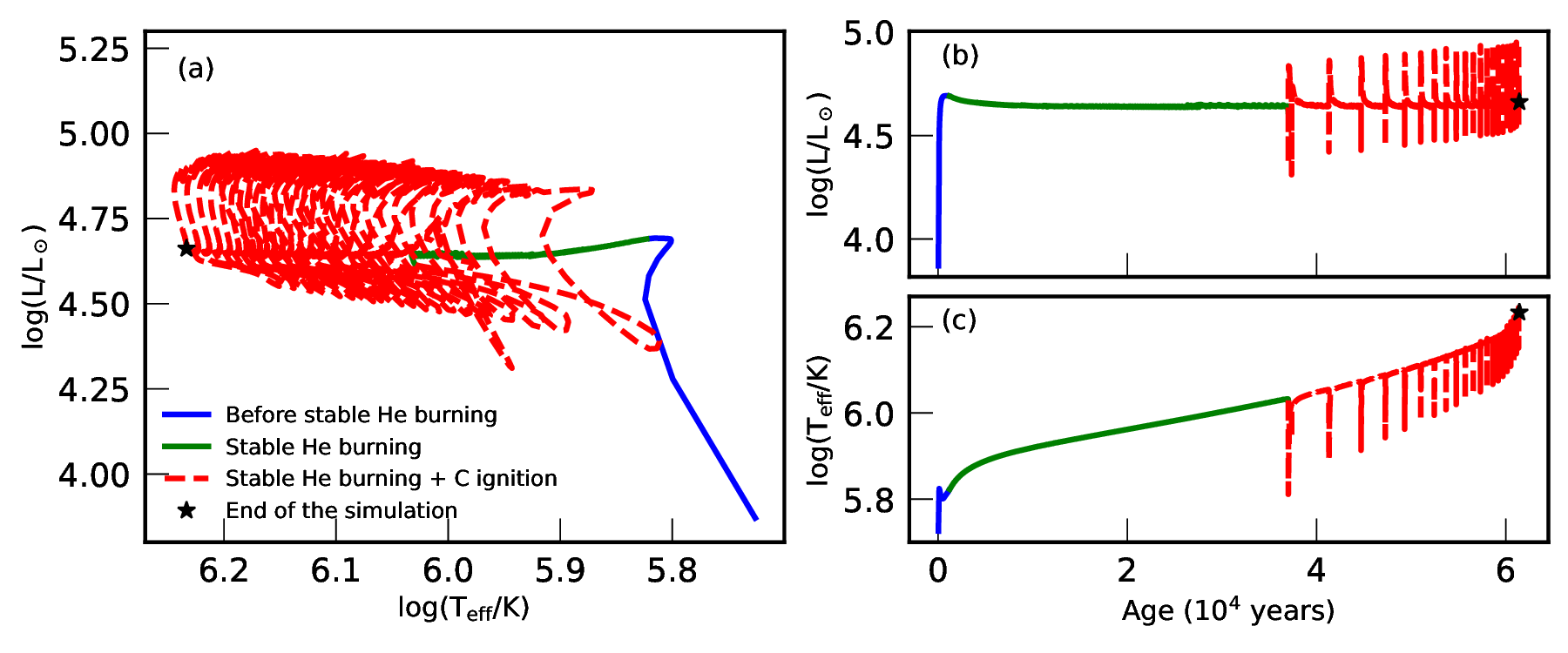}}
    \caption{Panel(a): HR diagram of the WD accretion process($\rm M_{WD} = 1.2 \,  M_{\odot}$, $\rm \dot{M}= 3 \times 10^{-6} \,  M_{\odot} \, yr^{-1}$). Three different colors represent different evolutionary stages. Panel(b): Evolution of luminosity over time. The three different lines have the same meanings as in Panel (a). Panel(c): Effective temperature as a function of time. The three different lines have the same meaning as in Panel (a). The periodic luminosity increase/decrease corresponds to successive temporal carbon burning (carbon shell flashes).}
    \label{fig1}
\end{figure}

Figure \ref{fig1}(a) presents the evolutionary track of a He-accreting WD on the Hertzsprung-Russell diagram (HRD). The entire evolutionary process can be divided into three stages: before stable helium burning, the stable helium-burning phase, and the helium/carbon alternately burning phase.

As accretion begins, the WD's surface temperature rises due to the release of gravitational energy from the accumulating helium-rich material. After approximately 100 years of accretion, helium burning enters a stable phase, and the WD's mass steadily increases. The steady mass increase of the WD mass is because the helium-rich layer burns stably if the accretion rate falls between $6 \times 10^{-7}$ and $3-4 \times 10^{-6}\, \rm M_{\odot}\, yr^{-1}$ \citep[e.g.][]{2017MNRAS.472.1593W, 2018ApJ...863..125K}. During the stable helium-burning phase, helium-rich material is converted into carbon and oxygen, while the evolutionary track on the HR diagram moves leftward. As carbon and oxygen accumulate at the bottom of the He layer, the temperature and density there increase, leading to an increase in the rate of nuclear reactions. Helium burning and mass accumulation persist for about 37,000 years in the stable helium-burning phase.

Once the $^{12}C + ^{12}C$ reaction rate surpasses the energy conduction rate and neutrino loss rate, carbon burning ignites. Carbon burning happens just below the helium-burning layer, triggering an inward-propagating carbon flame. Figure \ref{fig1}(b) and (c) clearly illustrate the timing of carbon ignition and the corresponding changes in luminosity and effective temperature. At the moment of carbon ignition, the luminosity surpasses the Eddington luminosity, triggering a super-Eddington wind and causing mass loss. During this period, helium burning is temporarily extinguished. After the carbon flame is quenched, the WD resumes stable helium burning and mass accumulation. At approximately 41,300 years, the second off-center carbon ignition occurs. Notably, silicon production becomes apparent from the second ignition onward. In the subsequent evolution, carbon ignition occurs repeatedly, with the ignition time intervals gradually decreasing. After each carbon-burning episode, a small peak of $^{28}\rm Si$ forms in the abundance profile. A total of 22 off-center carbon ignitions occurred throughout the evolution, with a total mass loss of about $7.44 \times 10^{-5}\rm M_{\odot}$. The simulation proceeded until an oxygen deflagration occurred at around 61,400 years, at which point we terminated the simulation. 

\begin{figure*}
    \centering
    \resizebox{0.7\hsize}{!}{\includegraphics{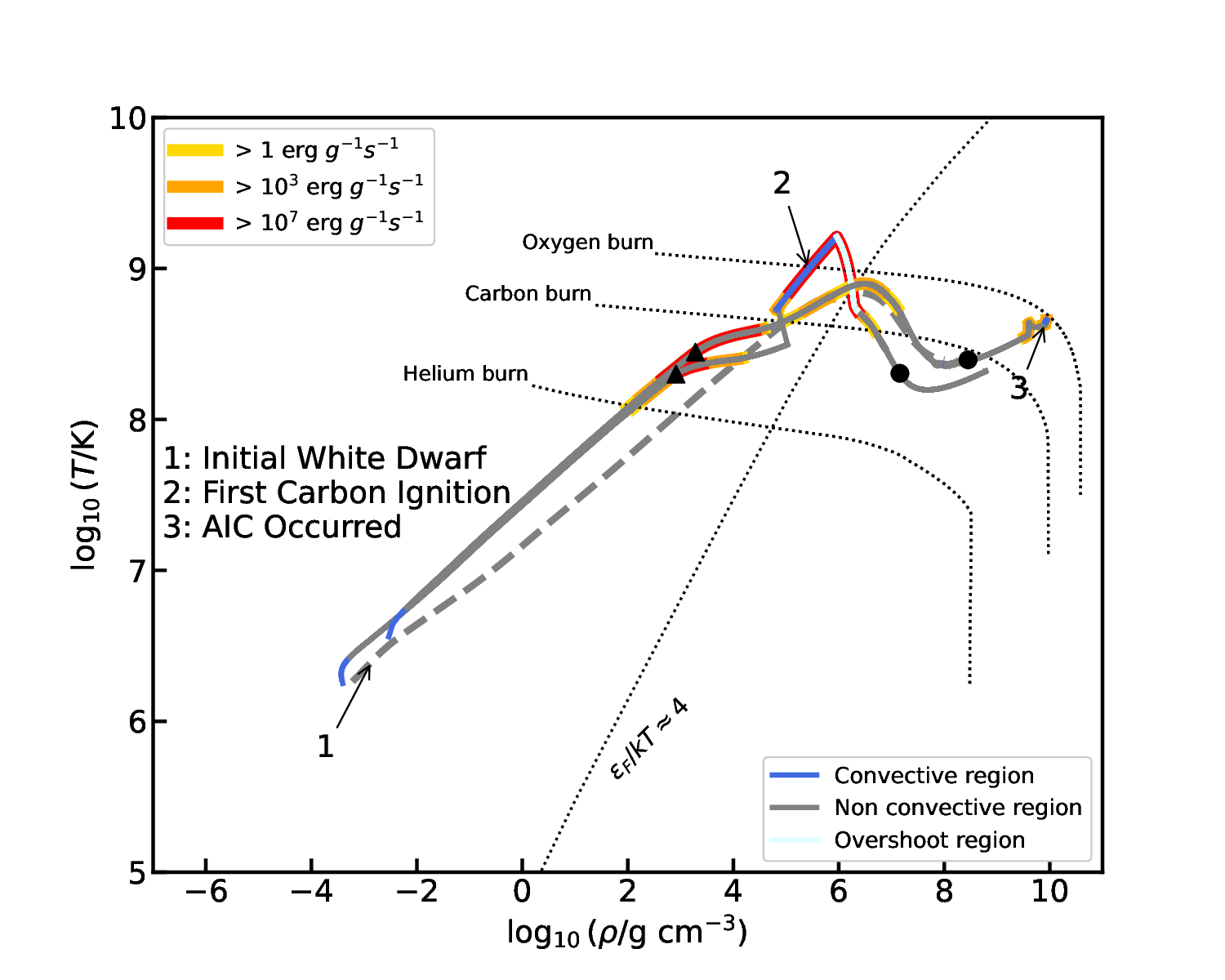}}
    \caption{Density-temperature profiles of the initial WD($\rm M_{WD} = 1.2 \,  M_{\odot}$), first carbon ignition, and when the AIC occurs, respectively. The accretion rate is assumed to be constant at $3 \times 10^{-6} \, \rm M_{\odot} \, yr^{-1}$ during the accretion process.} The black dotted lines represent $\varepsilon_{F}/kT \approx 4$ (where $\varepsilon_{F}$ is the Fermi energy), He, C, and O burning lines; the royal blue, grey, and light cyan curves represent the convective, non-convective and overshooting regions, respectively. The thick golden, orange, and red curves represent the regions where nuclear reaction rates are over $1$, ${10}^{3}$ and ${10}^{7}\,{\rm erg}/{\rm s}$ respectively. The black circle marks the boundary between the ONe core and the mantle, while the black triangle represents the boundary between the helium shell and the mantle.
    \label{fig2}
\end{figure*}

Figure \ref{fig2} presents the density-temperature profiles of the accreting WD at different evolutionary stages. The figure highlights the locations of the helium-burning shell and the core-mantle boundary (1.2$\rm M_{\sun}$). This region is where carbon is recurrently ignited and silicon is produced. The initial WD model has a luminosity of ${\log}(L/{\rm L_{\sun}}) \approx 3.87$, which is characteristic of a hot WD. As accretion proceeds, the core contracts, leading to increased central density and temperature. At the WD's surface, helium burning releases a substantial amount of energy, driving the formation of an outwardly extending convection zone and causing the expansion of the helium layer. The expansion of the helium envelope allows the WD to reach a state of equilibrium where radiative heat loss from the outer layers balances the energy generated by helium burning.
Off-center carbon ignition occurs approximately 37,000 years after the onset of accretion when the carbon-rich layer increases to about 0.1$\rm M_{\sun}$. The carbon flame reaches a temperature of roughly ${\rm log}(T/\rm K) \simeq 9.2$, significantly hotter than temperatures found in SAGB stars \citep[e.g.][]{2007A&A...476..893S, 2013ApJ...762....8D}. This difference in temperature results in different products from carbon burning. In SAGB stars, carbon burning primarily produces an ONe mixture without significant silicon production.

Figure \ref{fig3} displays the elemental abundance profiles of the WD at the moment of the first off-center carbon ignition and at the onset of AIC. The carbon flame reaches a temperature of ${\rm log}(T/\rm K) \simeq 9.2$. The initial carbon burning event primarily produces $^{20}\rm Ne$ and $^{24}\rm Mg$. At this stage, the production of $^{28}\rm Si$ is limited. The carbon flame itself is extremely thin, resulting in a steep temperature gradient, as seen in Figure \ref{fig3}(a). This flame propagates inwards and is eventually quenched at a mass coordinate of approximately 1.26$\rm M_{\odot}$ after 450 years. At this point, the WD has reached a mass of $1.31 \, \rm M_{\odot}$. As the WD continues to accrete mass, each subsequent carbon ignition increases the abundance of $^{28}\rm Si$, creating small peaks visible in Figure \ref{fig3}(b). Following the final ignition, the average silicon abundance in the mantle (1.2-1.378 $\rm M_{\odot}$) reaches 0.218.

Electron capture occurs when the timescale for compression becomes equal to the timescale for electron capture. Based on Figure 3 in \citet[][]{2017MNRAS.472.3390S}, significant electron capture by $^{24} \rm Mg$ begins when the core density reaches $\rm log(\rho_c/g \, cm^{-3}) = 9.6 $. As electron capture progresses in the core, the central temperature ($T_{\rm c}$) increases due to the exothermic nature of the electron capture reactions. The core is rapidly depleted of Mg, leading to a very steep gradient in the $^{24} \rm Mg$ fraction, as shown in Figure \ref{fig3}(b). When the core density reaches $\rm log(\rho_c/g , cm^{-3}) = 9.9 $, electron capture by $^{20} \rm Ne$ is triggered, causing a rapid increase in $T_{\rm c}$. The initially high mass fraction of $^{20} \rm Ne$ allows for a sustained period of $^{20} \rm Ne$ electron capture, which in turn triggers the $^{16} \rm O + ^{16} \rm O$ nuclear reaction. This oxygen burning further raises $T_{\rm c}$ to approximately ${\rm log}(T_{c}/\rm K) \approx 9.3$. \citet[][]{2015MNRAS.453.1910S} demonstrated that the energy released during $^{20} \rm Ne$ capture can trigger an oxygen deflagration wave when the core density reaches a critical value of $8.5 \times 10^{9} \, \rm g \, cm^{-3}$. \citet[][]{1992ApJ...396..649T} argued that if electron capture continues in the oxygen-burning remnants, objects igniting oxygen at such high densities are likely to collapse and form neutron stars. Note that this is subject to uncertainties in modeling the deflagration phase by multi-dimensional hydrodynamic simulations \citep{2016A&A...593A..72J,2019ApJ...886...22Z,2019PhRvL.123z2701K,2020ApJ...889...34L}, which is beyond the scope of this work. Here, we use the criterion that the central density exceeds the critical value $8.5 \times 10^{9} \, \rm g \, cm^{-3}$ at oxygen ignition as the condition of AIC. Our models with central densities at the time of oxygen ignition $\sim 8.511 \times 10^{9} \, \rm g \, cm^{-3}$ meet this criterion. Therefore, we conclude that the fate of our models will be AIC which leads to the formation of a neutron star.

\begin{figure*}[htbp]
    \centering
    \resizebox{0.9\hsize}{!}{\includegraphics{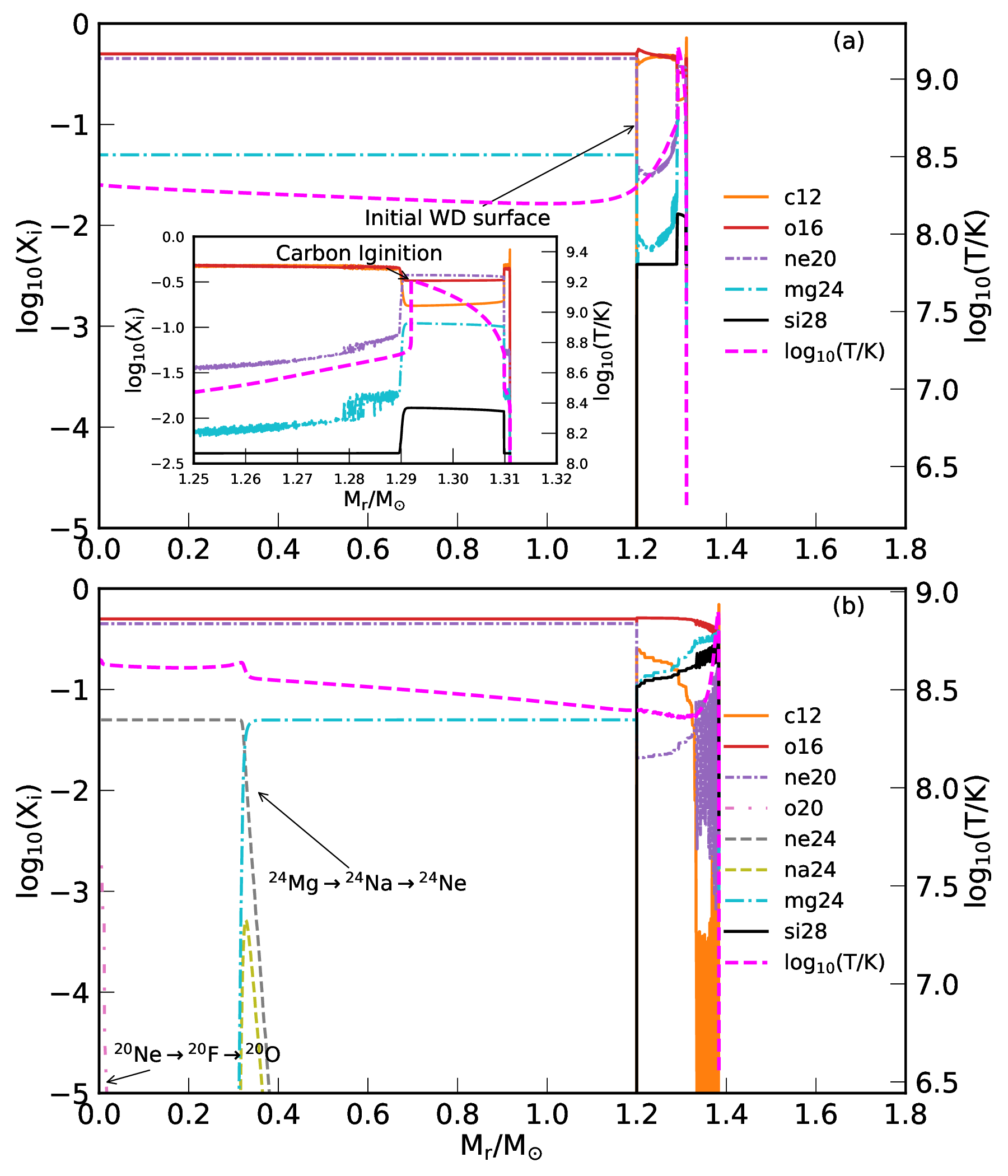}}
    \caption{Panel(a): Distribution of elemental abundances in WD at the onset of the first off-center carbon burning. The magenta dashed line shows the temperature profile of the WD. The black solid line represents the mass fraction of silicon, while the orange and red solid lines represent carbon and oxygen. Other colored dashed and dotted-dash lines represent the remaining elements. The inset shows the elemental abundance distribution between $\rm M_{r}=1.25-1.32\rm M_{\odot}$. Panel(b): Elemental abundance distribution in the WD at the moment of AIC. Labels are the same as in panel(a).The initial mass of the ONe WD is 1.2$\rm M_{\sun}$, and the accretion rate is $3 \times 10^{-6} \, \rm M_{\odot} \, yr^{-1}$.}
    \label{fig3}
\end{figure*}

\section{Discussion}\label{sec3}

\subsection{Comparison with previous work}\label{sec4.1}
\citet{1985A&A...150L..21S} investigated the merger evolution of two CO WDs. The merging process was modeled as a rapid mass transfer from the less massive white dwarf to the more massive one at a rate close to the Eddington accretion rate ($\rm \dot{M} \sim 10^{-5} M_\odot \, yr^{-1}$). It was found that the accretion ignited off-center carbon burning in a shell. The carbon burning front then propagated inward, passing through the central region, and eventually transformed the CO WD entirely into an ONe WD in a quiescent manner. This study is the first to propose the phenomenon of off-center carbon ignition during the WD accretion process. \citet{2016ApJ...821...28B}using the binary module of MESA, simulated the process of CO WD accreting material from a helium star. Their study revealed that, under specific orbital periods and accretion rates, the hot carbon ashes near the WD's outer edge can ignite, generating an inward-propagating carbon flame. This carbon flame propagates to the center, ultimately transforming the CO WD into an ONe WD. No production of $^{28}Si$ was observed in this process.
\citet{2019MNRAS.486.2977W} using the star module of MESA, simulated the process of CO white dwarf accreting helium-rich material. They discovered that the CO WD also undergoes off-center carbon ignition. Due to the high temperature of the carbon-burning flame, $\rm ^{20}Ne$ that appears is immediately burned into $\rm ^{28}Si$, converting the CO WD into an OSi WD. The main new finding in our paper is that helium-accreting ONe WDs experience off-center carbon ignition, where carbon can convert $\rm ^{20}Ne$ into $\rm ^{28}Si$. However, unlike the findings of \citet{2019MNRAS.486.2977W}, the carbon flame in our simulations extinguishes before reaching the ONe core, preventing the formation of an OSi core. Instead, this process results in a structure consisting of an ONe core surrounded by a silicon mantle(average silicon abundance in the mantle reaches 0.218).

\subsection{Urca process}\label{sec4.2}
The Urca process cooling can influence the central temperature evolution of accreting WDs. In the cooling process of WDs, two isotopes, $^{23}Na$ and $^{25}Mg$, play important roles. They participate in the Urca process through electron capture and $\beta$ decay, producing significant cooling effects and thus influencing the thermal evolution of the WDs.We have examined the conditions for the occurrence of the Urca process, and the results are similar to those of \citet{2017MNRAS.472.3390S}. The density of oxygen ignition is about $\rm log\,(\rho_{c}/g \, cm^{-3}) \approx 9.95$, very close to the value without Urca process cooling. This suggests that the Urca process cooling cannot alter the final fate of the accreting ONe WDs, which will still undergo AIC.

\subsection{Different initial masses and accretion rates}\label{sec4.3}
The initial mass of the WD may have a significant effect on the final fate of the helium accretion WDs. Figure~\ref{fig5}(a) shows the final mass fraction of $^{28}\rm Si$ for different initial WD masses. We find that under the same accretion rate, WDs with different masses eventually accumulate almost the same amount of $^{28}\rm Si$. The production of $^{28}\rm Si$ is insensitive to the initial mass of the WD.

The accretion rate may also affect the final results of helium accretion WDs. We compare the difference in $^{28}\rm Si$ fraction resulting from carbon ignition at different accretion rates in Figure~\ref{fig5}(b). At a lower accretion rate, the carbon layer is massive when it ignites compared to the high accretion rate, resulting in a higher silicon fraction. At the same time, the first ignition of the carbon is so intense that the convection zone extends outward in the high accretion rate case, so there will be a thick layer of silicon. We also found that in the $4\times 10^{-6}\, \rm M_{\odot}\, yr^{-1}$ and  $5\times 10^{-6}\, \rm M_{\odot}\, yr^{-1}$ cases, the carbon flame can penetrate the initial surface of the WD. The reason for this phenomenon is that, at these two accretion rates, a significant compositional gradient exists between the ONe core and the CO material produced by helium burning during the helium-burning phase. Carbon flames are convection-bound, and abundance gradients vanish within the flame as it propagates toward the core/shell interface. Additionally, if the flame extinguishes at the surface of the ONe core, a significant portion of the CO burning products will be ONeMg, leading to a decrease in the gradient near the core surface. During long-term evolution, the temperature near the surface of the core decreases, and the radiative temperature gradient gradually shifts from negative to 0. At a certain point, it surpasses the sum of the adiabatic temperature gradient and the compositional gradient. Consequently, under the Ledoux criterion employed in our simulations, convective instability arises near the core surface, leading to the diffusion of elements. However, this result is subject to current numerical uncertainties.

\subsection{Different initial WD temperatures}\label{sec4.4}
If the initial temperature of the model is too low, the WD will not be in thermal equilibrium. It will undergo helium and carbon flashes at the beginning of accretion. Whether or not a helium flash and a carbon flash occur at the beginning of accretion will affect the accretion products and may affect the conclusions of the study. As shown in Figure \ref{fig6}, we compared two WD models($1.2\rm M_{\odot}$, $3\times 10^{-6}\, \rm M_{\odot}\, yr^{-1}$) with different initial temperatures, specifically $\rm logT_{eff} = 5.73$ and $\rm logT_{eff} = 5.53$. The convection criterion for all models adopted the Ledoux criterion. We found that the model with $\rm logT_{eff} = 5.73$ did not experience significant initial helium or carbon flashes, yet the final $^{28}\rm Si$ mass fraction remains consistent with the lower temperature model. Quantitatively, the final average $^{28}\rm Si$ abundance in the mantle ($\rm M_{r}=1.2-1.378 \rm M_{\odot}$) is 0.211 for the $\rm logT_{eff} = 5.53$ model and 0.218 for the $\rm logT_{eff} = 5.73$ model. This indicates that whether the WD initially undergoes helium and carbon flashes or not has minimal impact on the final $^{28}\rm Si$ mass fraction.

\begin{figure}
\centering
    \resizebox{0.8\hsize}{!}{\includegraphics{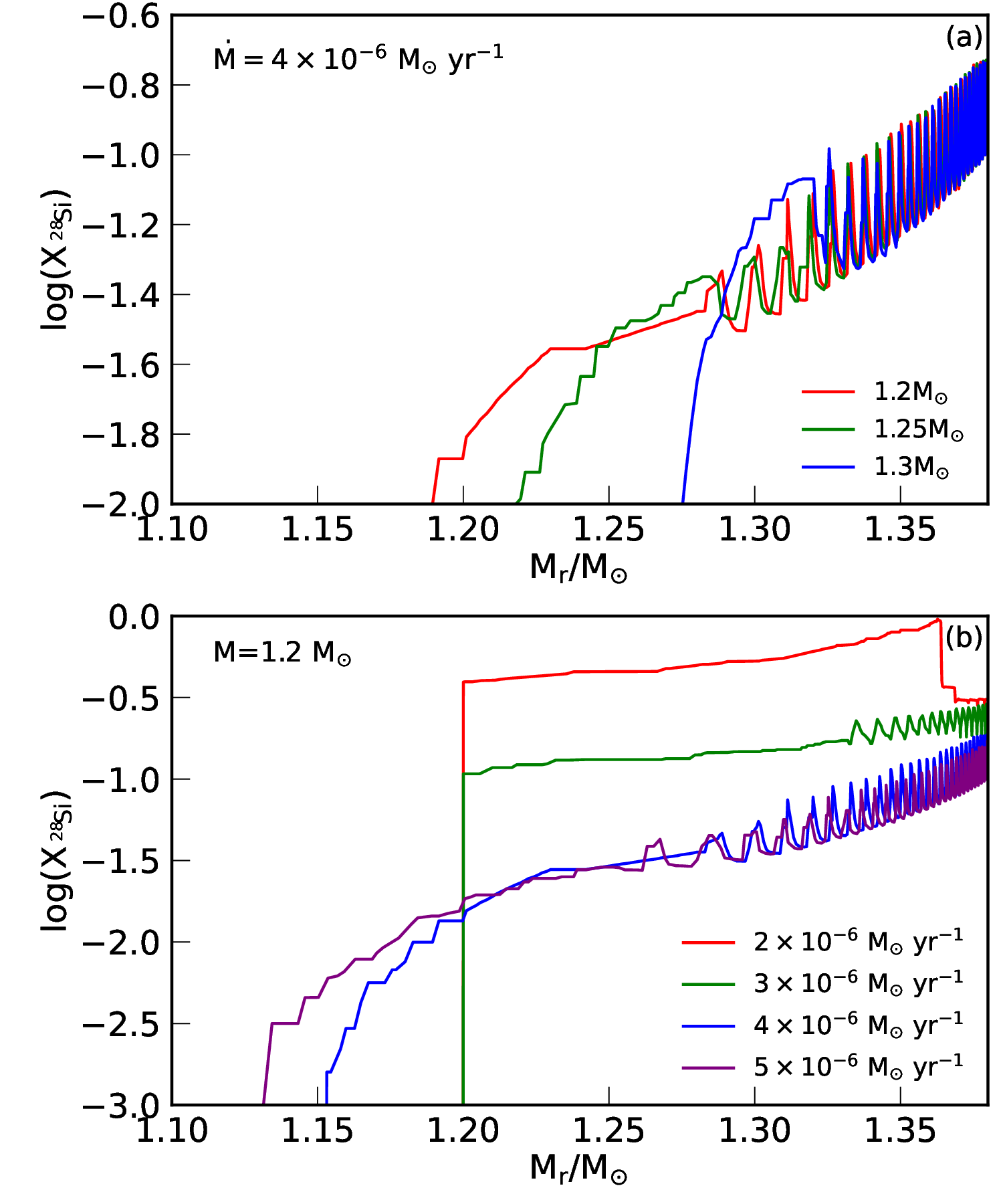}}
    \caption{Panel(a): Silicon mass fractions in WDs with varying initial masses. These simulations were conducted with a constant accretion rate of $4 \times 10^{-6} \rm \, M_{\odot} \, yr^{-1}$. Panel(b): Silicon mass fractions in WDs with a uniform initial mass of $1.2\rm M_{\odot}$, but with varying accretion rates.}
    \label{fig5}
\end{figure}

\begin{figure}
    \centering
    \resizebox{1\hsize}{!}{\includegraphics{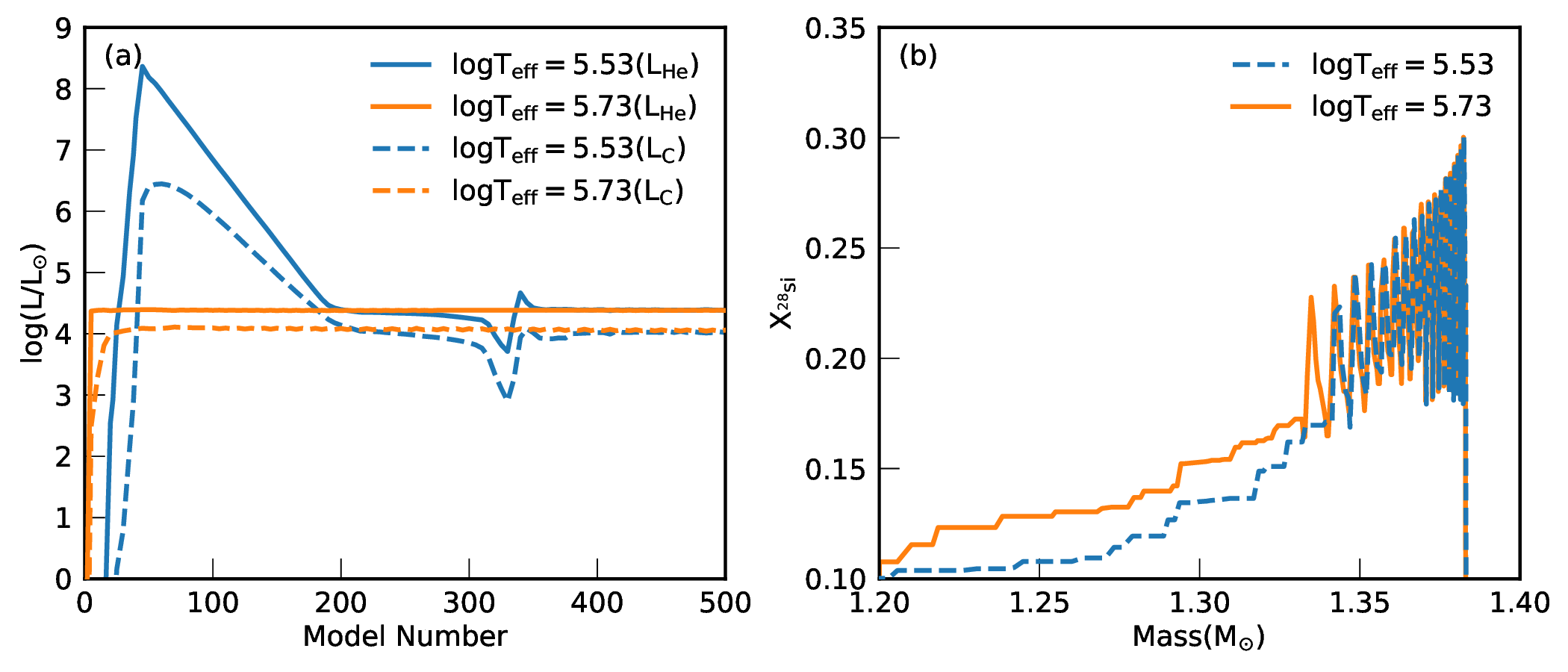}}
    \caption{Panel(a): Initial helium and carbon luminosities for $1.2\rm M_{\odot}$ WDs at $3 \times 10^{-6} \rm \, M_{\odot} \, yr^{-1}$ with different initial effective temperatures at the beginning of evolution. Panel(b): $^{28}\rm Si$ abundances for $1.2\rm M_{\odot}$ WDs at $3 \times 10^{-6} \rm \, M_{\odot} \, yr^{-1}$ with different initial effective temperatures at the end of the simulation.}
    \label{fig6}
\end{figure}

\section{Summary}\label{sec4}

By employing MESA, we explored the long-term evolution of He-accreting ONe WDs by considering different initial WD masses ($\rm M^{i}_{WD} = 1.2, 1.25, \ and\  1.3 M_{\odot}$) and accretion rates ($2\times 10^{-6} - 5\times 10^{-6} \rm M_{\odot} \, yr^{-1}$) in our simulations. We find that off-center carbon burning leads to a direct conversion of carbon in the outer layer of WDs to silicon. The mass fraction of silicon is sensitive to the mass accretion rate of WDs, but insensitive to the initial mass of WDs. Lower accretion rates lead to higher silicon mass fractions. If the central density exceeds the critical density of $8.5 \times 10^{9} , \rm g , cm^{-3}$, oxygen deflagration will occur. Electron capture will continue in the oxygen-burning remnants, and WDs that ignite oxygen at such high densities are likely to form NSs by AIC.

The most distinctive feature of this type of AIC event is the potential presence of a significant silicon component in its spectrum. Unfortunately, no AIC events have been observed to date. If future transient surveys discover AIC events, our model has the potential to be confirmed through spectroscopic observations.

\section{Acknowledgments}

We thank Keiichi Maeda, Takashi J. Moriya, and Zhengwei Liu for their discussions and help. This study is supported by the National Natural Science Foundation of China (Nos 12225304, 12288102, 12393811, 12090040/12090043, 12273105, 12473032), the National Key R\&D Program of China (No. 2021YFA1600404), the Yunnan Revitalization Talent Support Program -- Young Talent project, the Western Light Project of CAS (No. XBZG-ZDSYS-202117), the science research grant from the China Manned Space Project (No. CMS-CSST-2021-A12), the Frontier Scientific Research Program of Deep Space Exploration Laboratory (No. 2022-QYKYJH-ZYTS-016), the International Centre of Supernovae, Yunnan Key Laboratory (No. 202302AN360001), the Youth Innovation Promotion Association CAS (No. 2021058), and the Yunnan Fundamental Research Project (Nos 202201BC070003, 202401AV070006, 202201AW070011).

\bibliographystyle{aasjournal}
\bibliography{work1}

\end{document}